\documentstyle[aps]{revtex}
\input epsf
\draft
\begin{document}

\twocolumn[
\hsize\textwidth\columnwidth\hsize\csname@twocolumnfalse\endcsname
\title{Percolation and jamming in random sequential adsorption of linear
segments on square lattice}

\author{
Grzegorz Kondrat$^1$ and Andrzej P\c{e}kalski$^2$}

\address
{Institute of Theoretical Physics, University of Wroc{\l}aw,
pl. M. Borna 9, 50-204 Wroc{\l}aw, Poland\\
 email addresses : $^1$ gkon@ift.uni.wroc.pl, $^2$
 apekal@ift.uni.wroc.pl}
\maketitle

\begin{abstract}

We present the results of study of random sequential adsorption
of linear segments (needles) on sites of a square lattice.
We show that the percolation threshold is a nonmonotonic function of
the length of the adsorbed needle, showing a minimum for a certain length of the needles,
 while the jamming threshold
decreases to a constant with a power law. The ratio of the two thresholds is
also nonmonotonic and it remains constant only in a restricted
range of the needles length. We determine the values of the
correlation length exponent for percolation, jamming and their
ratio.


\end{abstract} ]

\section {INTRODUCTION}
The problem of percolation is an old one \cite{ham} but still new
results appear and some unsolved questions remain \cite{stau}. In
general site percolation is defined on a $d$-dimensional lattice
where each site can be either occupied with the probability $c$
or empty with the probability $1-c$. Neighboring occupied sites
form a cluster. If it is so large that it reaches the two opposite
edges of the lattice, e.g. top and bottom, the cluster is said to
be percolating. The lowest concentration of occupied sites for
which there is a percolating (or spanning) cluster for an infinite
lattice is called the percolation threshold $c_p$ \cite{stau}.

Another realization of the percolation problem is random
sequential adsorption (RSA), in which objects (point particles,
segments, rectangles etc.) are put on randomly chosen sites and
the objects do not move \cite{evans}. It is also possible to
consider RSA in a continuum \cite{ziff}.

Jamming is a problem related to RSA percolation \cite{evans}. Again
objects are placed randomly on the lattice sites until a concentration
$c_j$ is reached, where there is no room on the lattice for the
next object. For point like particles $c_j$ = 1, but for spatially
extended entities $c_j < 1$. Continuum models of jamming also
exist \cite{evans}.

The RSA models
irreversible dissociation \cite{balazs} and binding of large
ligands to polymer chains \cite{boucher}. Another area of
applicability is the deposition of large molecules on solid
surfaces, like proteins \cite{ramsden} or macromolecules on
biological membranes \cite{finegold}.  The isotropic-nematic
transition in the hard rods like polymers has been studied first
by Flory \cite{flory} and later e.g. in \cite{binder}. Spatial
organization of needles into a well organized nematic phase is
however a
different problem, not considered here. General forms of
percolation models have a wide range of applications - from
chemisorption, spatially disordered systems, porous materials, car
parking and ecology \cite{evans}, to separating the good and bad
people at the entrance to Hades \cite{domb}.
For overview of percolation, jamming and related problems see  \cite{evans}.

In a recently published paper \cite{nicolas} Vanderwalle et al.
studied the relation between the two transitions - percolation and
jamming. They used two kinds of objects - linear segments of
length 2 to 10 and square blocks. They have found that the ratio
of the two threshold concentrations $c_p$ and $c_j$ is constant
$c_p$/$c_j \approx 0.62$, regardless of the length of the needle.

In the present paper we extend the study of Vanderwalle et al. to
larger lattices and longer objects (we consider only linear
segments). In particular we shall check the claim that
the $c_p/c_j$ ratio does not depend on the length of the needles.

\section{THE MODEL}
We consider a square lattice of size $L \times L$. On the sites of
the lattice we put randomly linear segments (needles) of a given
length $a$, with the constraint that the needles cannot cross each
other, although they may touch themselves. We used hard boundary conditions,
i.e., the needles may touch the edge of the lattice but they cannot stick out of it -
each needle must lay totally inside the lattice. Adopting open boundary
conditions does not affect the results.

 To achieve simulation
efficiency, our algorithm of deposition needles consists of two
parts designed for two different regimes. Firstly when the
current concentration of the needles is small, we chose randomly,
from a uniform distribution, the orientation (vertical or
horizontal) and position of the upper left end of the needle to
be inserted. If there is enough space on the lattice, the needle
is deposited, if not we pass to the next try. After a certain number of adsorption
trials we switch to the other regime where the dense
routine is applied. A list of all empty sites and orientations
still available is made. From that list a site is randomly
chosen. We determine the direction of the needle and check
whether the needle can be put there. In any case the site is
removed from the list. The process is continued till
the last item on the list. Such organization saves time, since we
avoid to insert needles into densely packed regions.

 A cluster is defined as
a group of sites linked by the needles. If there is an uninterrupted
path between the top and the bottom of the lattice, the cluster is
said to be percolating or spanning, and the concentration of
occupied sites defines the percolation threshold $c_p$. The concentration
at which no more needles could be put on the lattice without
violating the constraint determines the jamming threshold
$c_j$.

We have considered lattices of sizes $L$ = 30, 100, 300, 1000, 2500
and needles of length $a = 1 \,\, ..\,\,  2000$. On the smallest lattices
only smaller needles were located. Averaging was done over 100
independent runs. We have checked that averaging over 1000 runs
did not reduce the error (mean standard deviation - $\sigma$) in a
marked way.

\section{RESULTS}
Our main results of the simulations are presented in Figure 1, where the percolation
and jamming thresholds, as well as their ratio, are plotted
against the length of the needles ($a=1\,\, .. \,\, 45$). These data are
obtained for lattice size $L=2500$. As convergence and error analysis shows
 (see below) we can safely accept them as the asymptotic ($L\rightarrow\infty$) values.

\begin{figure}
\centerline{\epsfxsize=7cm \epsfbox{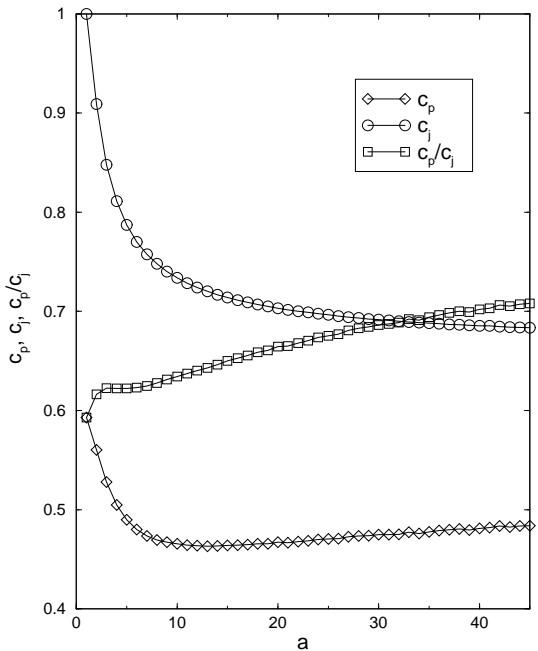}} \caption{Thresholds
for percolation $c_p$, jamming $c_j$ and their ratio $c_p/c_j$
 {\it versus} needles' length, $a$. Lattice size $L$ = 2500.
 Averaged over 100 samples.}
\end{figure}

The percolation threshold for $a = 1 \,\, .. \,\, 13$ diminishes,
then it begins to grow linearly with the slope 0.00071. The minimum value
$c_{p\, min}=0.463$ is reached for $a=13$. As seen in
Figure 2, the $\sigma$ increases with the size of the needles starting from
$0.001$  ($a = 1$) up to $0.008$ ($a = 45$).
\begin{figure}
\centerline{\epsfxsize=7cm \epsfbox{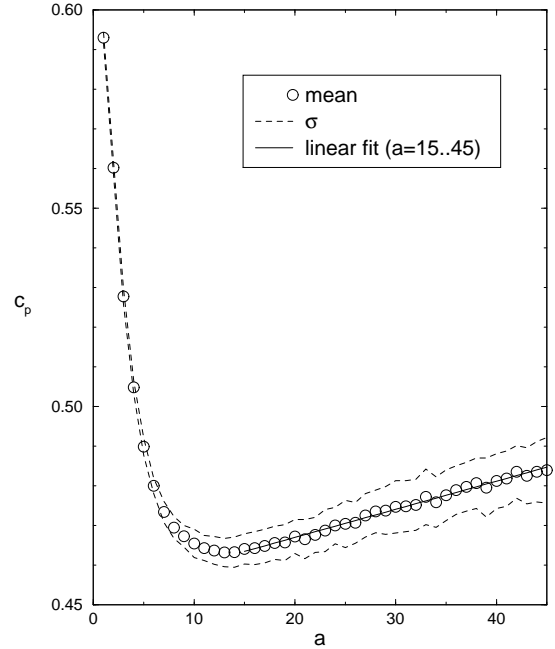}}
\caption{Percolation threshold  $c_p$ {\it versus} needles'
length $a$. $L$ = 2500, 100 runs.  Short needles $a = 1\, ..\,
45$.}
\end{figure}
\begin{figure}
 \centerline{\epsfxsize=7cm \epsfbox{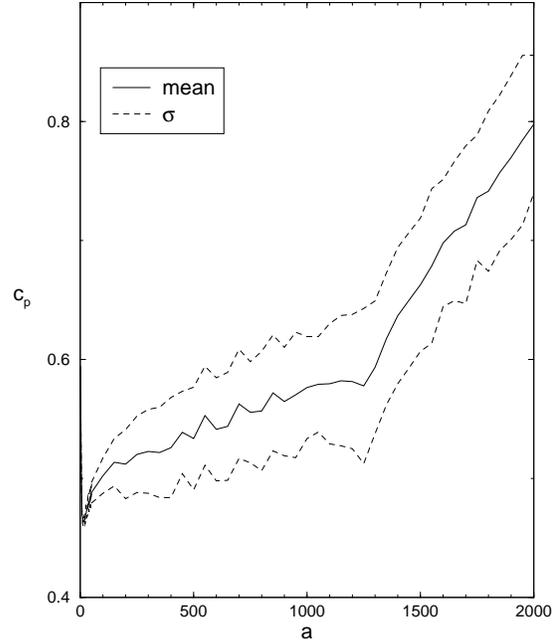}}
\caption{Percolation threshold  $c_p$ {\it versus} needles'
length $a$. $L$ = 2500, 100 runs.  Long needles $a$ =
1\,..\,2000.}
\end{figure}
The increase of the percolation threshold for longer $a$ is
however quite clear. The appearance of this novel and unexpected
feature is connected with the condition that the needles may
touch themselves but they cannot cross. In the simulations where
the restriction has been lifted we observed no minimum but a
monotonic decrease.  In the model considered here  the needles
have the tendency to align in parallel not only with respect to
the edges of the lattice but also to themselves (see Figures 4
and 5), hence the needles form compact clusters.

\begin{figure}
 \centerline{\epsfxsize=7cm \epsfbox{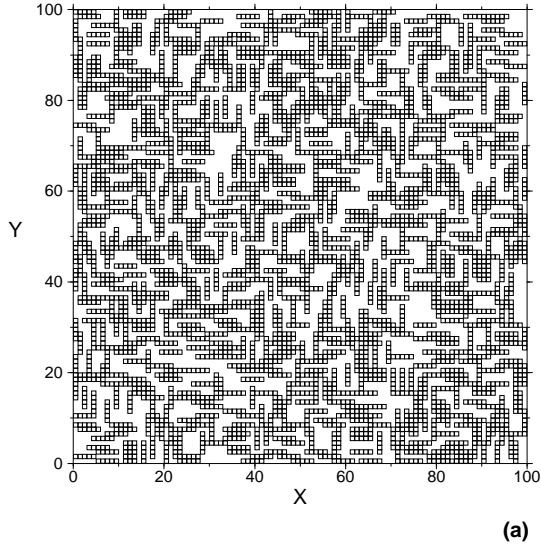}}
\caption{Snapshot of a spatial distribution of needles at the
percolation threshold for $L$ = 100, $a$ = 5.}
\end{figure}
\begin{figure}
 \centerline{\epsfxsize=7cm \epsfbox{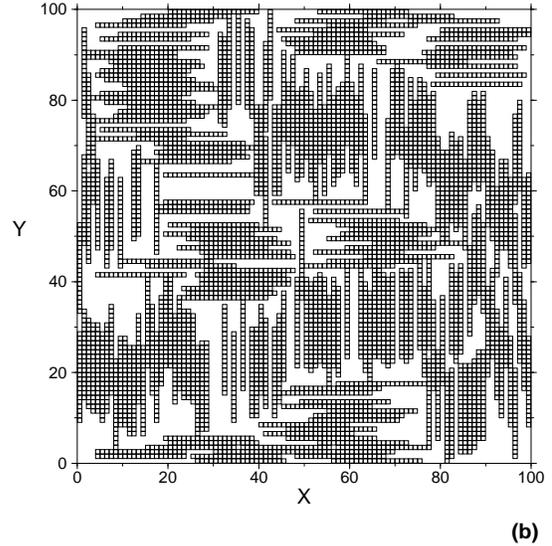}}
\caption{Snapshot of a spatial distribution of needles at the
percolation threshold for $L$ = 100,  $a$ = 20.}
\end{figure}
 In case of
horizontally oriented needles in order to move e.g. two steps
down, two needles of length $a$ are needed. The longer are the
needles the higher is
 the percentage of occupied sites necessary for passing these two steps. The increase of $c_p(a)$ is to a certain degree
compensated by vertically oriented needles, which however also form clusters,
thus offering many equivalent ways for percolation.
 Further simulations
for much longer needles
indicate continuous increase in $c_p$, although at a slower rate - see Figure 3.\\
The jamming thresholds obtained from the simulations have much
smaller error than that for percolation and even for $a=45$ it is
below $0.002$. Values of $c_j$, as a function of $a$, decrease
according to a power law (very good fit for all $a\geq 5$)
approaching the asymptotic value $c^*_j = 0.66\pm 0.01$ (see
Figure 6):
\begin{equation}
c_j = c^*_j + 0.44 \cdot a^{-0.77}.
\end{equation}
\begin{figure}
 \centerline{\epsfxsize=7cm \epsfbox{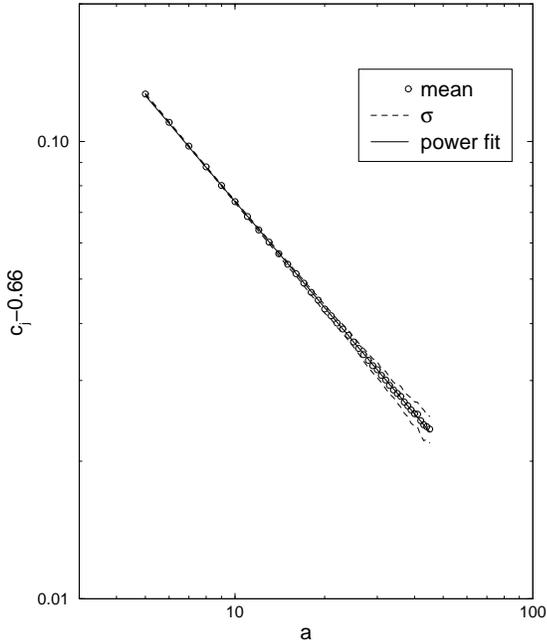}}
\caption{Jamming threshold $c_j$ {\it versus} needles' length $a$
on a log-log plot. $L$ = 2500, 100 runs. $a = 5\,..\,45$.}
\end{figure}
The uncertainty of the exponent derived from the graph analysis
equals $0.02$. Clearly this behaviour differs essentially from bare
power law postulated in \cite{ziff} :
\begin{equation}
c_j \sim a^{-0.2},
\end{equation}
for the continuum (off - lattice) case of RSA of randomly oriented and highly
anisotropic (length to wide)
rectangles. Their $a$ coincides with our
length of needles $a$. In the discrete case we did not observe the maximum of $c_j$ at $a=2$ reported in \cite{ziff}. The reason is that on the lattice the number of possible orientations of the needles is restricted to $z/2$ (where $z$ is the coordination number of the lattice) in contrast to the continuum case. It is interesting that the asymptotic concentration for jamming  (for $a\rightarrow\infty$) is $0$ off lattice  and it remains finite in the discrete case.

Other interesting quantity in our model is the ratio $c_p/c_j$ as a function of
 $a$ (see Figure 7).
\begin{figure}
 \centerline{\epsfxsize=7cm \epsfbox{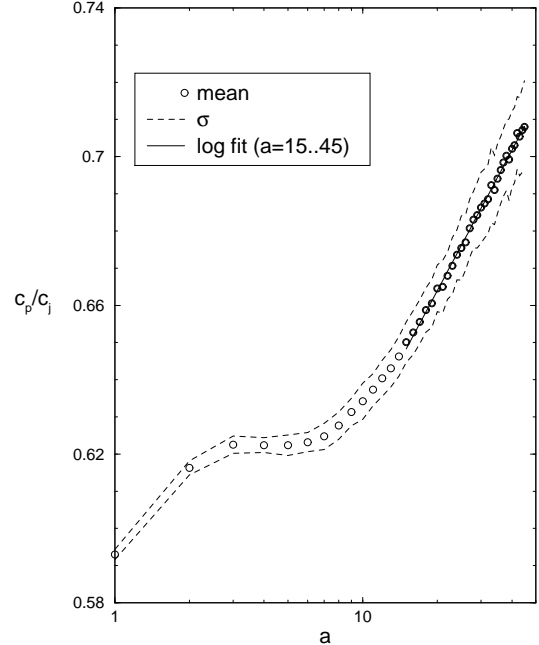}}
\caption{Percolation to jamming thresholds ratio $c_p/c_j$ {\it
versus} needles' length $a$. $L$ = 2500, 100 runs. Logarithmic fit
for $a = 15\,..\,45$.}
\end{figure}
 It grows  for $a = 1 \,\, .. \,\, 3$, then it
stabilizes till $a = 7$ and then it grows again. The plateau value of
$c_p/c_j\approx 0.62$ , the constant found in \cite{nicolas}. The growth for
longer needles ($15\leq a \leq 45$) could be fitted by a logarithmic
dependence\begin{equation}
c_p/c_j\sim 0.50 + 0.13 \cdot \log \, a.
\end{equation}
\begin{figure}
 \centerline{\epsfxsize=55mm \epsfbox{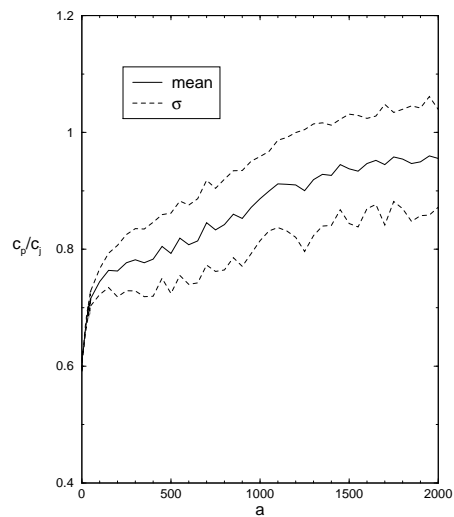}}
\caption{Percolation to jamming thresholds ratio $c_p/c_j$ {\it
versus} needles' length $a$. $L$ = 2500, 100 runs. $a =
1\,..\,2000$.}
\end{figure}
Further simulations for longer needles (see Figure 8) support our
claim of monotonic increase in $c_p/c_j$ over wide range of $a$
(even up to $a=2000$). We may conclude therefore that the
universality claimed in \cite{nicolas} holds only in a rather
restricted range of $a\in[3,7]$. As a matter of fact, the value
of $c_p/c_j$ for $a>7$ shown in Table 1 of \cite{nicolas} is
greater than those for $a\leq 7$ but the authors attribute it to
the finite size effects. This is however most probably just the
beginning of the growth of $c_p/c_j$.

We analyzed the dependence of the obtained thresholds on the
lattice size $L$ and needles' length $a$ focusing on convergence.
It appeared that for the ratio $a/L<1/3$ the values of $c_p$ and
$c_j$ do not vary much with increasing $L$ (keeping $a$ constant)
- see Figures 9 and 10.
\begin{figure}
 \centerline{\epsfxsize=7cm \epsfbox{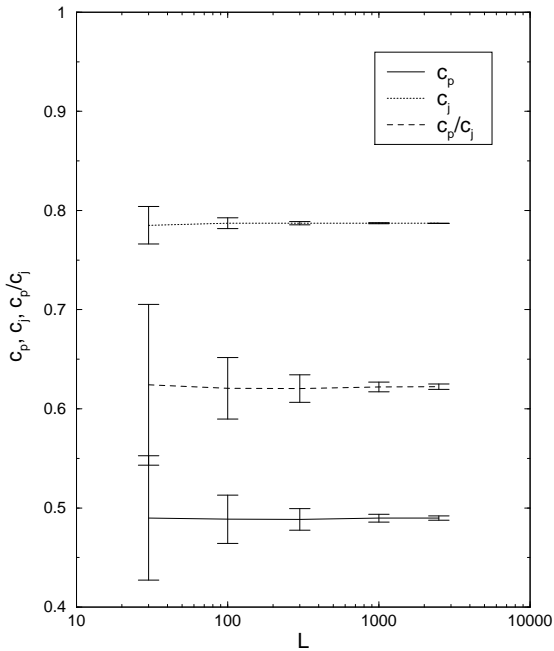}}
\caption{Convergence analysis of percolation, $c_p$, jamming
thresholds, $c_j$, and their ratio $c_p/c_j$, {\it versus}
lattice size $L$. 100 runs, $a$ = 5.}
\end{figure}
\begin{figure}
 \centerline{\epsfxsize=7cm \epsfbox{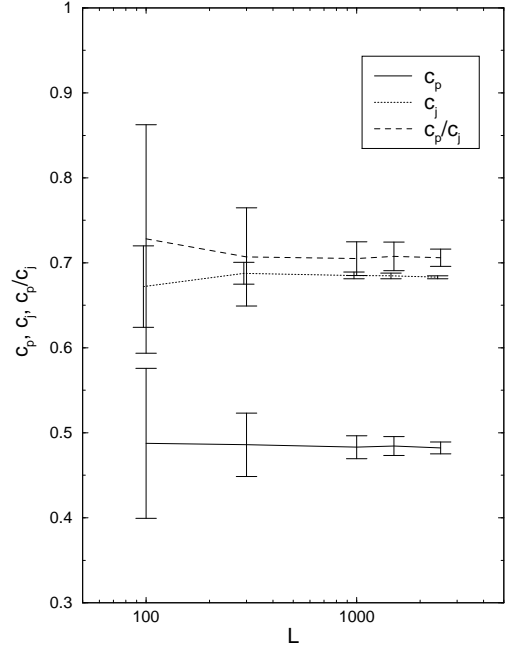}}
\caption{Convergence analysis of percolation, $c_p$, jamming
thresholds, $c_j$, and their ratio $c_p/c_j$, {\it versus}
lattice size $L$. 100 runs, $a$ = 45.}
\end{figure}
The error bars (here $\sigma$) however decrease rapidly with $L$,
whilst the difference of the thresholds for different lattice
sizes is much smaller than the appropriate error. Thus it is safe
to take the values of the thresholds from the simulations with
$L=2500$ as the asymptotic (exact) ones.

The finite size effects can clearly be seen in Figure 3, where
$c_p$ is drawn against $a=1 \,\, .. \,\, 2000$ for $L=2500$. At
$a=L/2$ we can notice sharp change in the slope of the function
$c_p(a)$.

Consider now the dependence of $\sigma$ of $c_p, c_j, c_p/c_j$ on
the lattice size. $\sigma$ is analogous to the quantity $\Delta$
defining in \cite{nicolas} the sharpness of the transition
(non-percolating to percolating or non-jammed to jammed). Here
however the power law approach to the asymptotic value
$p(\infty)-p(L) \sim L^{-1/\nu}$  (cf. formula (3) in
\cite{nicolas} ) does not hold.
\begin{figure}
 \centerline{\epsfxsize=7cm \epsfbox{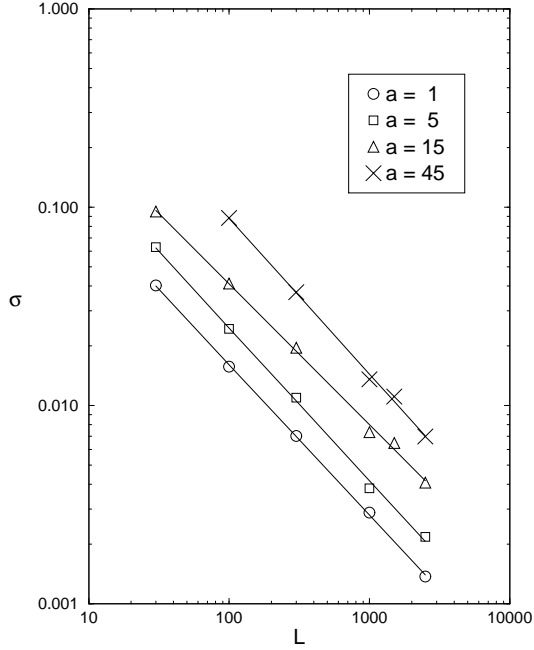}}
\caption{Deviation analysis. $\sigma$ {\it versus} lattice size
$L$ for several values of the needles' length.  Percolation.}
\end{figure}
\begin{figure}
 \centerline{\epsfxsize=7cm \epsfbox{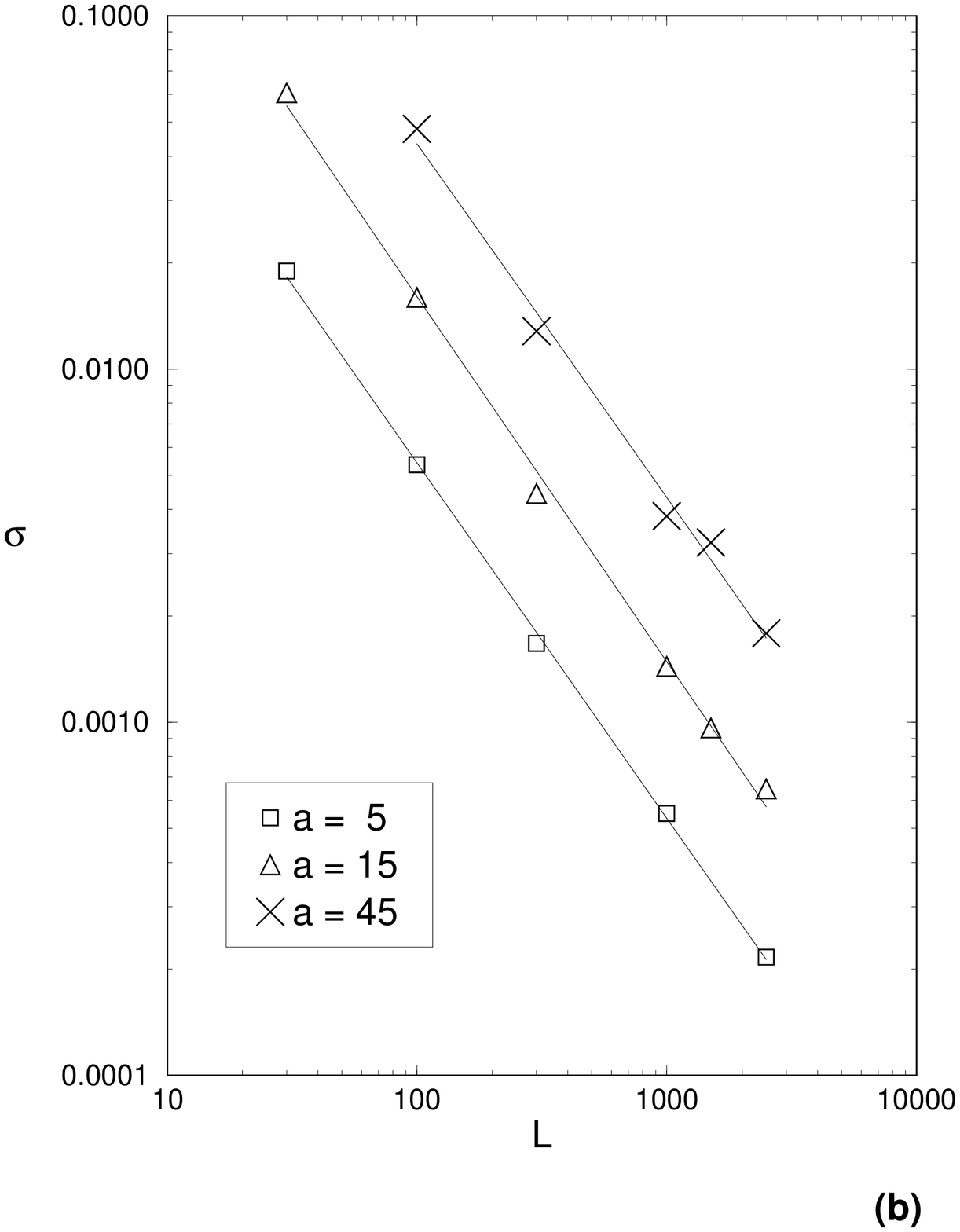}}
\caption{Deviation analysis. $\sigma$ {\it versus} lattice size
$L$ for several values of the needles' length.  Jamming.}
\end{figure}
\begin{figure}
 \centerline{\epsfxsize=7cm \epsfbox{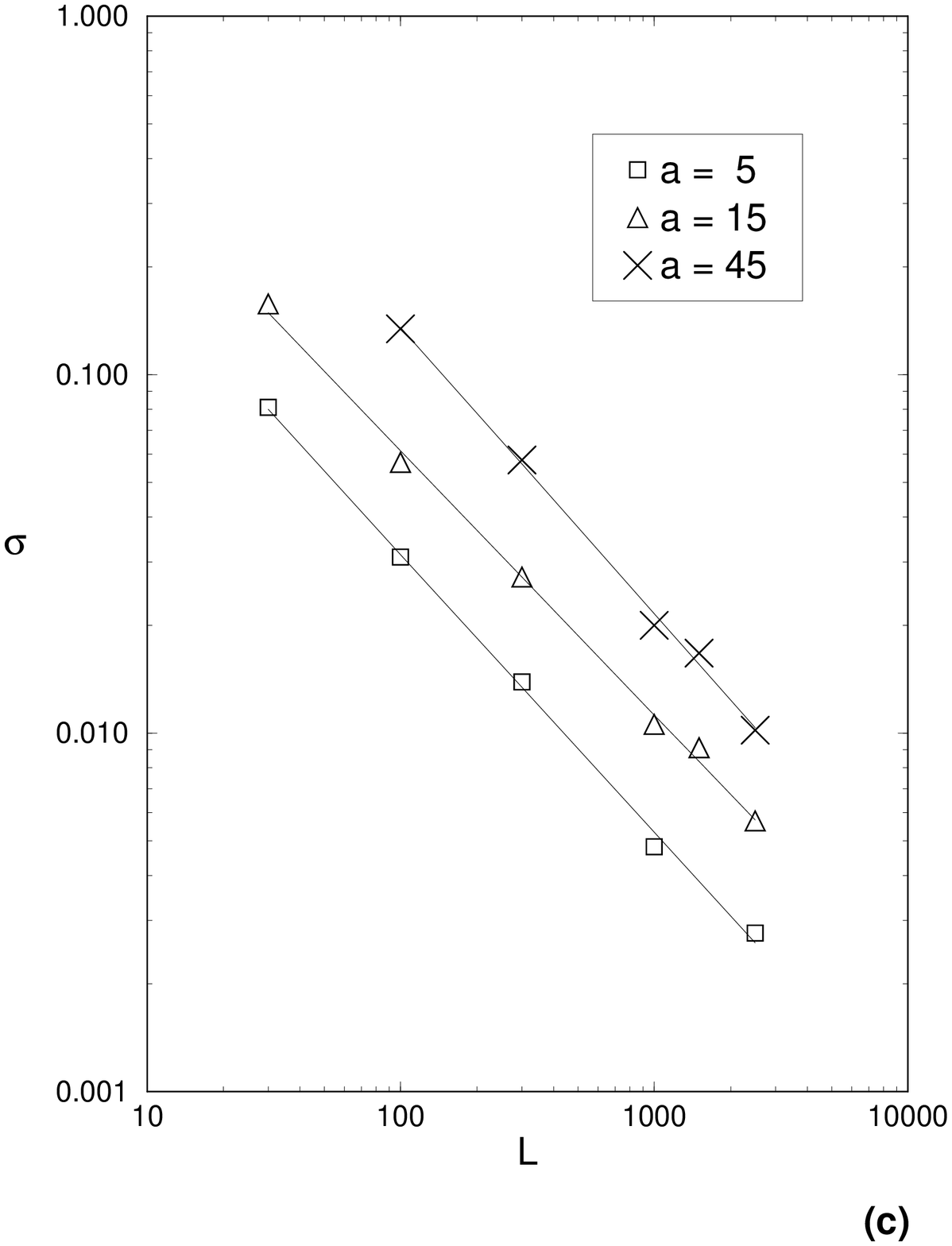}}
\caption{Deviation analysis. $\sigma$ {\it versus} lattice size
$L$ for several values of the needles' length.  Percolation to
jamming ratio.}
\end{figure}

We have found (see Figures 11 - 13) that the $\sigma$ for
percolation ($\Delta_p$), jamming ($\Delta_j$) and the $c_p/c_j$
ratio ($\Delta_r$) decrease with the lattice size according to
the power laws
\begin{eqnarray}
  \Delta_p \sim L^{-1/\nu_p}, \,\,\,\,\,\, 1/\nu_p = 0.75 \pm 0.05,\nonumber\\
  \Delta_j \sim L^{-1/\nu_j}, \,\,\,\,\,\, 1/\nu_j = 1.00 \pm 0.05,\\
  \Delta_r \sim L^{-1/\nu_r}, \,\,\,\,\,\, 1/\nu_r = 0.77 \pm 0.05.\nonumber
\end{eqnarray}
Here $\nu$ corresponds to the correlation length exponent \cite{stau}
\begin{equation}
\xi \sim \mid c - c_p \mid^{-\nu}.
\end{equation}
These values are, within the error bars, the same for all
$a = 1 \,\, .. \,\, 45$ and agree with those found by Vanderwalle et al
\cite{nicolas}. Also Nakamura \cite{nakamura} found $\nu_j = 1.0
\pm 0.1$ for RSA of square blocks. It seems therefore that the exponents $\nu$
are good candidates for  universal quantities.

\begin{figure}
 \centerline{\epsfxsize=7cm \epsfbox{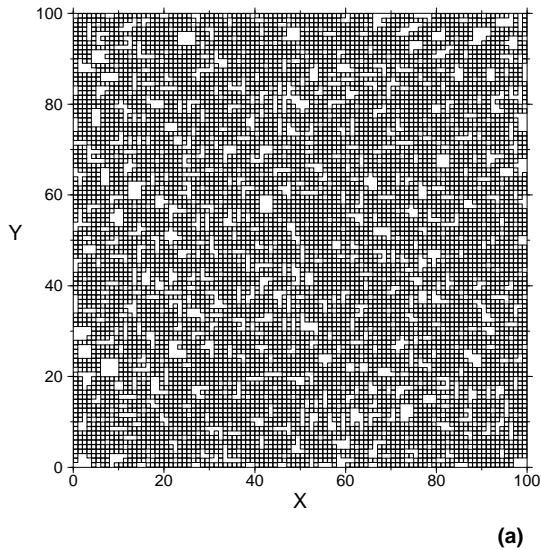}}
\caption{Snapshot of a spatial distribution of needles at the
jamming threshold for $L$ = 100, $a$ = 5.}
\end{figure}
\begin{figure}
 \centerline{\epsfxsize=7cm \epsfbox{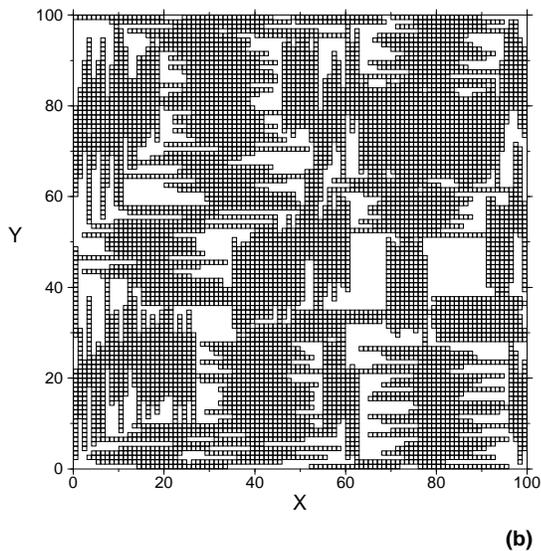}}
\caption{Snapshot of a spatial distribution of needles at the
jamming threshold for $L$ = 100,  $a$ = 20.}
\end{figure}
Examples of spatial arrangements of shorter ($a=5$) and longer
($a=20$) needles on a lattice $100 \times 100$ are shown in
Figures 4 and 5 (percolation) and in Figures 14 and 15 (jamming).
Analysis based on examination of different runs shows some
regularity in the needles distribution -- we have found that the
needles near the edges have the tendency to stick along the
borders. Longer needles, for obvious reasons, form clusters of
parallel alignment, as was already observed in \cite{nicolas}.

\section{CONCLUSIONS}
We have performed extensive simulations of RSA using linear segments
of size $a = 1 \,\, .. \,\, 45$ on square lattice sites. We have found
that the percolation threshold is a nonmonotonic function of $a$, having a
minimum due to parallel orientation of the needles, at $a$ = 13,
while the jamming threshold decreases to a non-zero constant
with $a$ as a power law. The
ratio of the two thresholds is nonmonotonic too - after initial
growth it stabilizes for some values of $a$, and then it grows
logarithmically. Whether the asymptotic value is equal to one or below it
is an interesting question. To answer it unequivocally is unfortunately
beyond our computing power. The values of the correlation length exponent
$\nu$, for percolation, jamming thresholds and the ratio of the
two, do not depend on the length of the needles and they are,
within the error bars, equal to those found elsewhere
\cite{nicolas}\cite{nakamura} for deposition of needles, rectangles or
squares.\\[5mm]
\acknowledgements

We are grateful to M.Droz, J.O.  Indekeu, Z. Koza and N. Vandewalle for helpful comments.

\end{document}